\documentclass[authoryear, 12pt,1p]{elsarticle}
\usepackage{lipsum}
\makeatletter
\def\ps@pprintTitle{%
 \let\@oddhead\@empty
 \let\@evenhead\@empty
 \def\@oddfoot{}%
 \let\@evenfoot\@oddfoot}
\makeatother

\usepackage{graphicx}
\usepackage{color}
\usepackage{amsfonts}
\usepackage{amsmath}
\usepackage{amsthm}
\usepackage{array}
\usepackage{verbatim}
\usepackage{mathabx}
\usepackage{natbib}
\usepackage{sidecap} 
\usepackage{float}

\begin{document}

\begin{frontmatter}

\title{Simultaneous Reduction of Two Common Autocalibration Errors in GRAPPA 
EPI Time Series Data}

\author[ucb]{D. Sheltraw \corref{cor2}}
\ead{sheltraw@berkeley.edu}

\author[siem]{V. Deshpande}

\author[ucb]{M. Trumpis}

\author[ucb]{B. Inglis \corref{cor1}}
\ead{binglis@berkeley.edu}

\cortext[cor1]{Corresponding author}
\cortext[cor2]{Principal corresponding author}

\address[ucb]{Henry H. Wheeler Jr. Brain Imaging Center, Helen Wills 
Neuroscience Institute, University of California, Berkeley, CA, 94720, USA}
\address[siem]{MR R\&D, Siemens Healthcare, Neuroscience Imaging Center, UCSF
1 Irving St., Room A-C 109 San Francisco, CA 94122}

\begin{abstract} The GRAPPA (GeneRalized Autocalibrating Partially Parallel 
Acquisitions) method of parallel MRI makes use of an autocalibration scan 
(ACS) to determine a set of synthesis coefficients to be used in the image
reconstruction. For EPI time series the ACS data is usually acquired once prior
to the time series. In this case the interleaved $R$-shot EPI trajectory, where
$R$ is the GRAPPA reduction factor, offers advantages which we justify from a 
theoretical and experimental perspective. Unfortunately, interleaved $R$-shot 
ACS can be corrupted due to perturbations to the signal (such as direct and
indirect motion effects) occurring between the shots, and these perturbations
may lead to artifacts in GRAPPA-reconstructed images. Consequently we also 
present a method of acquiring interleaved ACS data in a manner which can reduce
the effects of inter-shot signal perturbations. This method makes use of the 
phase correction data, conveniently a part of many standard EPI sequences, to 
assess the signal perturbations between the segments of $R$-shot EPI ACS scans.
The phase correction scans serve as navigator echoes, or more accurately a
perturbation-sensitive signal, to which a root-mean-square deviation 
perturbation metric is applied for the determination of the best available 
complete ACS data set among multiple complete sets of ACS data acquired prior 
to the EPI time series. This best set (assumed to be that with the smallest 
valued perturbation metric) is used in the GRAPPA autocalibration algorithm, 
thereby permitting considerable improvement in both image quality and temporal 
signal-to-noise ratio of the subsequent EPI time series at the expense of a 
small increase in overall acquisition time. 
\end{abstract}

\end{frontmatter}

\section{Introduction}
\label{intro}

Parallel accelerated MRI \citep{SMASH, autoSMASH, SENSE, griswold1} increases 
image acquisition speed by using the spatial sensitivity of each receiver 
coil in an array of receiver coils, in addition to the spatial encoding 
provided by the applied linear magnetic field gradients used in conventional 
(non-accelerated) imaging. Relative to conventional imaging this additional 
spatial information allows one to reduce the number of acquired phase-encoded 
(PE) lines of data, while maintaining a desired digital resolution and 
field-of-view (FOV), thereby accelerating the data acquisition. This increase 
in the image acquisition speed is usually stated in terms of the reduction 
factor $R$ which is defined, for fixed nominal digital image resolution and 
FOV, as the ratio of the number of acquired PE lines for a conventional scan 
to the number of acquired PE lines for a parallel accelerated imaging scan. 

\begin{figure}[H]
  \centering
  \includegraphics[scale = 1.0, trim = 70 375 350 575, clip, viewport = 0 375 450 575]{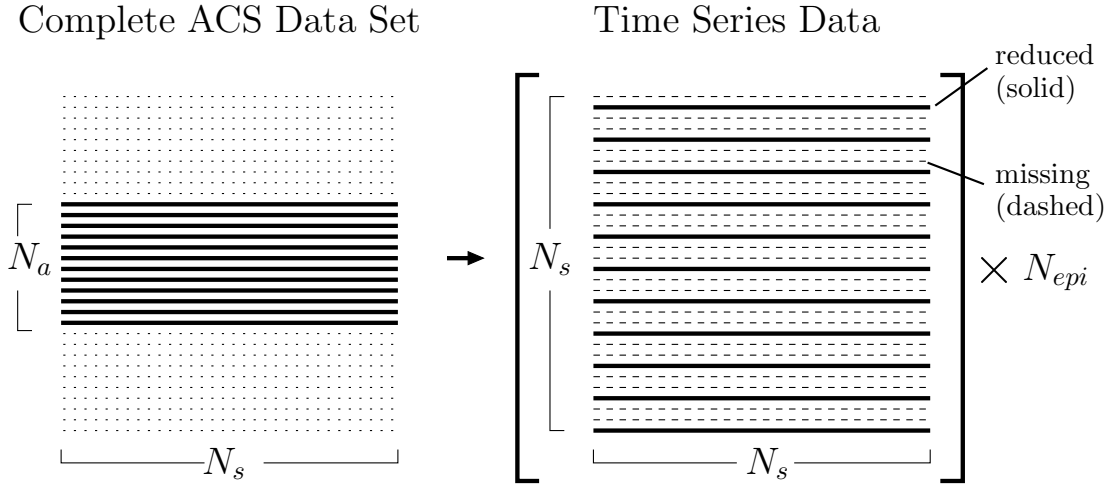}
  \caption{1-shot ACS acquisition: This figure illustrates the ACS and time 
  series k-space data associated with a single 2D image slice for the $R=3$ 
  case. The bold arrow indicates temporal order of acquisition and the 
  ${\bigtimes}$ symbol denotes multiple acquisitions of the data contained 
  within the vertical square brackets. The solid horizontal lines in k-space 
  denote the acquired data. A {\it complete set of ACS data} consists of $N_a$ 
  PE lines which are collected during 1-shot EPI acquisition. In the time 
  series data the {\it reduced set} and the {\it missing set} of PE lines are
  respectively denoted by the solid and dashed lines. There are $N_s$ PE lines 
  in the nominal set of PE lines (nominal matrix size: $N_s \times N_s$).}
 \label{single_shot_acs}
\end{figure}

\begin{figure}[H]
  \centering
  \includegraphics[scale = 0.8, trim = 0 170 400 575 , clip, viewport = 0 170 500 575]{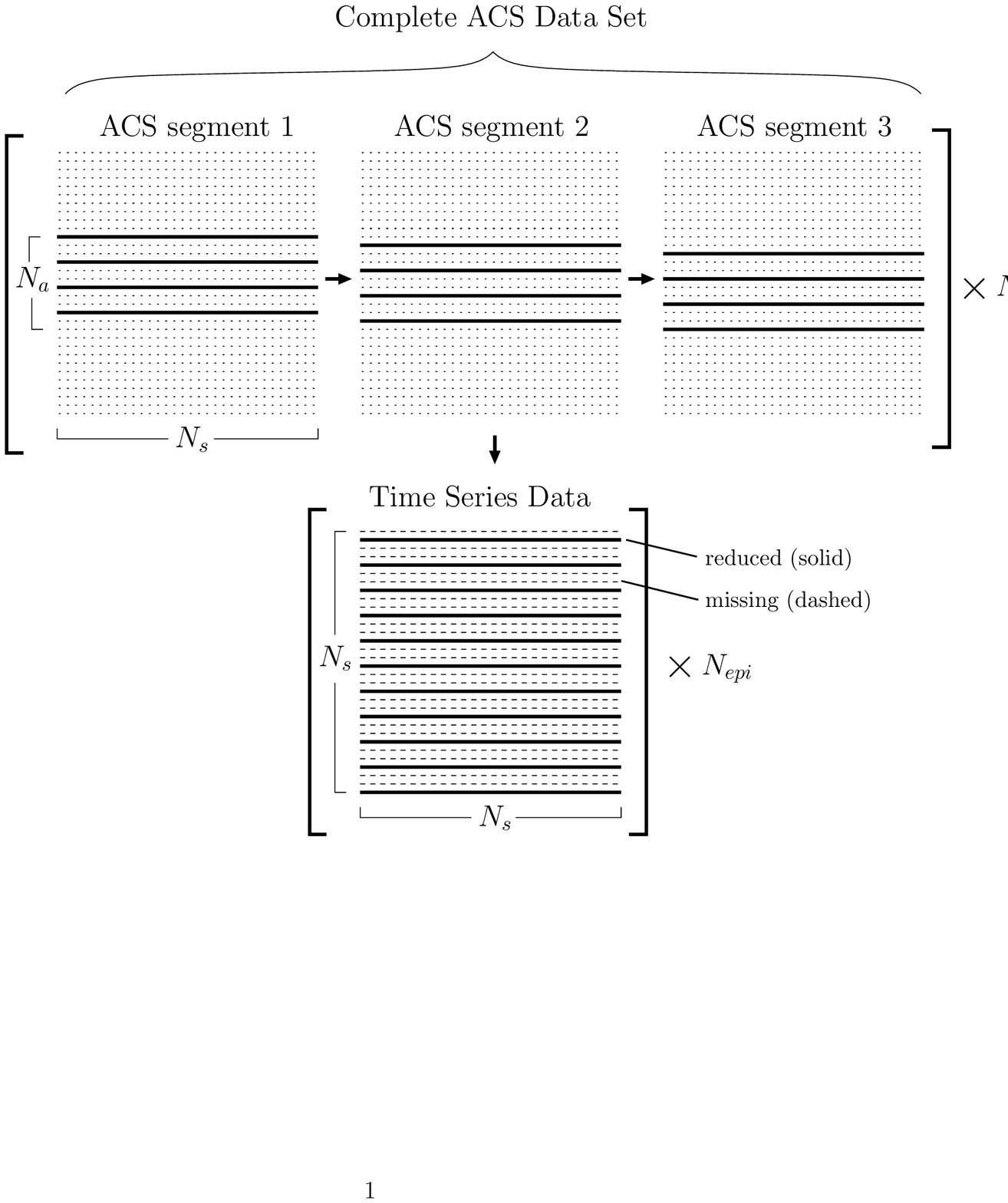}
  \caption{R-shot segmented ACS acquisition: This figure illustrates the ACS 
  and time series k-space data associated with a single 2D image slice for the 
  $R=3$ case. The bold arrows indicate temporal order of acquisition and the 
  $\bigtimes$ symbol denotes multiple acquisitions of the data contained 
  within the vertical square brackets. The solid horizontal lines in k-space 
  denote the acquired data. A {\it complete set of ACS data} consists of $N_a$ 
  PE lines which are collected during $R$ shots of a segmented EPI acquisition.
  In the time series data the {\it reduced set} and the {\it missing set} of 
  PE lines are respectively denoted by the solid and dashed lines. There are 
  $N_s$ PE lines in the nominal set of PE lines (nominal matrix size: $N_s 
  \times N_s$).}
 \label{pe_terms}
\end{figure}

For the sake of clarity we make some definitions before continuing with this
introduction. In keeping with terminology initiated in Griswold et al. 
\citep{griswold1} we will refer to the data sampled during accelerated imaging 
as the {\it reduced set} of PE lines and we will refer to the PE lines which 
are absent from the parallel accelerated imaging data set, but present in a 
conventional data set, as the {\it missing set} of PE lines. In addition we 
will refer to the union of the reduced set and the missing set as the {\it 
nominal set} of PE lines - equivalent to the set of PE lines acquired during  
conventional imaging - since it is the sampling characteristics of this set 
which determine the nominal digital image resolution and FOV. The right-hand 
side of Figure \ref{single_shot_acs} and the lower part of Figure 
\ref{pe_terms} illustrate these definitions as they pertain to an EPI 
time series.

The GRAPPA method of parallel imaging \citep{griswold1} has been of great 
interest since, unlike other parallel imaging methods such as SENSE 
(Sensitivity Encoding) \citep{SENSE}, it does not require explicit 
knowledge of the receive fields for each element of the receiver array. 
Instead, the GRAPPA method uses the data from the receiver array in an 
autocalibration procedure which estimates a set of {\it synthesis 
coefficients} used to synthesize the missing set of PE lines from the reduced 
set of PE lines over the set of receiver coils. Central to the autocalibration 
procedure is the acquisition of autocalibration scan (ACS) data consisting of 
a subset of the nominal set of PE lines. This subset, comprised of PE lines 
near the center of k-space, will be referred to as the {\it complete set of 
ACS data}. This complete ACS data set is used to calculate the synthesis 
coefficients over the set of receiver array channels. The left-hand side of 
Figure \ref{single_shot_acs} and the upper part Figure \ref{pe_terms} 
illustrate these definitions as they pertain to $N_a$ PE lines of a complete 
ACS data set acquired prior to an EPI time series for $1$-shot and
$R$-shot interleaved ACS, respectively.

For GRAPPA EPI time series the ACS data is usually collected once prior to the 
acquisition of the time series data \citep{Schmied}. Both $1$-shot and $R$-shot 
interleaved $k$-space trajectories \citep{Schmied} have been used to acquire a 
complete set of ACS data. Figures \ref{single_shot_acs} and \ref{pe_terms} 
depict the k-space trajectories for $1$-shot and $R$-shot interleaved segmented 
acquisition respectively, of complete ACS data sets for the $R=3$ case. The 
justification for the use of the $R$-shot as opposed to the $1$-shot 
trajectory has not been clearly established in the literature, nor has a 
robust method been established for dealing with the potential consequences of 
inter-shot signal perturbations (such as motion) during acquisition of a 
complete $R$-shot ACS data set. 

The intent of this paper is two-fold: (1) To justify the use of the $R$-shot 
interleaved EPI trajectory instead of the $1$-shot trajectory from a 
theoretical and experimental perspective when in the presence of significant
main magnetic field inhomogeneity, and (2) To demonstrate a method of 
acquiring the $R$-shot interleaved EPI ACS data which minimizes potential 
artifacts due to signal perturbations occurring between the $R$ shots. We 
will not attempt to answer the question of under what circumstances (main 
field strength, susceptibility differences, FOV choice, and reduction factor) 
the artifact from motion may be greater than that from trajectory 
incompatibility. Our hope is that with refinement of the method presented 
here this trade-off will simply not be in question.

Although in this paper we focus on the direct effects of head motion - because 
we can control it to some extent - we note that ACS data acquired in a 
segmented manner can potentially be contaminated by any perturbation to the 
data that occurs on a timescale smaller than that required to sample a 
complete ACS data set. Such perturbations may include, for example, motion 
effects such as $B_0$ changes due to subject chest motion, magnetic 
susceptibility changes due to motion, perturbations to the time series 
steady-state when through-slice-plane motion occurs and signal spiking due 
to electrostatic discharge between cables.  

We note that motion can perturb GRAPPA EPI time series data by two means: (1) 
Motion during the ACS data acquisition can lead to an inaccurate estimation 
of the synthesis coefficients, thereby leading to artifacts in all images of 
the time series along with a potential degradation of tSNR (temporal 
signal-to-noise ratio); and (2) Motion between the time of the ACS data 
acquisition and the time at which any particular image of the time series is 
acquired may make the synthesis coefficients inappropriate since the imaged 
object may have moved into regions where insufficient signal existed during 
the ACS data acquisition - an incomplete spatial sampling. 

Cheng \citep{Cheng} has investigated the use of TGRAPPA \citep{BKGJ}, a method 
not yet commercially available, to provide new ACS lines for each volume of an 
EPI time series. This has the benefit of providing motion-uncontaminated ACS 
throughout the time series, but comes at the expense of reducing tSNR in the 
absence of significant subject motion, compared to the performance of GRAPPA.
Thus, for fMRI applications, it is to be expected that GRAPPA would be 
preferable to TGRAPPA, provided GRAPPA can be made more robust to motion.

One of the methods in this paper involves assessing motion contamination
of ACS data by the use of a navigator echo \citep{RLE} and a perturbation
metric. Other investigators have studied methods by which to assess and reduce 
motion contamination of MRI data albeit in applications other than GRAPPA 
EPI time series. For example, Kim and Hu have used navigator echoes to limit 
the effects of motion in fMRI studies using the FLASH sequence \citep{XHSGK} 
and the interleaved EPI sequence \citep{KHAU}. 2D navigators have been used in 
conjunction with read-out segmented EPI \citep{Heidemann_Porter} \citep{Nguyen} 
for diffusion imaging, the navigator being used to determine whether a 
particular diffusion weighted image should be re-acquired due to motion 
contamination. Holdsworth et. al. \citep{HSNB} have also used a k-space entropy 
metric to assess motion corruption in read-out segmented EPI for 
diffusion-weighted imaging. Law et. al. \citep{Law_Glover} used a sliding 
window approach to update the coil sensitivity maps for the TSENSE (adaptive 
SENSE incorporating temporal filtering) \citep{Kellman_McVeigh} method. Here we 
extend the application of navigator echoes to assessing motion between ACS 
segments, a direct extension of previous ideas, albeit with a new application.

The method presented here makes use of multiple $R$-shot interleaved ACS EPI 
data sets and phase correction echoes, which are already part of most 
commercial EPI sequences, to assess the motion between ACS EPI segments and 
produce a complete set of ACS EPI data that is minimally corrupted by motion.
The phase correction echoes are therefore doing double-duty for they will be
used in their usual capacity to eliminate Nyquist ghosting and they will be
used as navigator echoes to assess motion. This complete set of ACS segments,
assessed to be minimally contaminated by motion, is then used to estimate 
the GRAPPA synthesis coefficients and hence synthesize the missing PE lines 
for the entire time series. With this redundant ACS scheme, $R$-shot 
accelerated EPI time series can potentially be reconstructed with less 
artifact and greater tSNR in the presence of motion. This suggests
improvements for fMRI applications which we evaluate with a simple measurement 
of tSNR.

\section{Theory}
\label{background}

In the GRAPPA method \citep{griswold1} the missing set of PE lines, of a 
nominal set with inter-sample distance $\Delta k_y = 1/{\rm FOV}$ in the PE 
direction, are synthesized according to the following equation:
\begin{eqnarray}
 S_n(m + r, x) 
 = \sum_{n'=1}^{N_c} \sum_{m' = -b}^{b} 
 S_n'(m + m' R, x) W_{nn'm'r}
 \label{sig0b}
\end{eqnarray}
where $R$ is the reduction factor, $m = \ldots -2R, -R, 0, R, 2R, \ldots$
enumerates the reduced set of PE lines, $r = 1, \ldots, R-1$ is a PE offset 
from a PE line of the reduced set to a neighboring PE line of the missing set, 
$N_c$ is the number of coil elements in the receiver array, $W_{nn'm'r} 
(\tau_y)$ are the synthesis coefficients and $b$ fixes a finite number of 
local PE lines of the reduced set to be used in the synthesis of the missing 
PE lines. Equation \ref{sig0b} is used in conjunction with the sampled ACS 
lines to perform an autocalibration step by which one obtains the synthesis 
coefficients through a fitting algorithm. Once the coefficients $W_{nn'm'r}$ 
are determined then Equation \ref{sig0b} is used to synthesize the missing set 
of PE lines from the reduced set of PE lines for each receiver coil. The 
images for each coil are then reconstructed by a simple FFT (Fast Fourier
Transform) of the nominal set of PE lines, followed by a square-root of a 
sum-of-squares combination \citep{Roemer, Larsson} of the individual coil 
images to produce the final image.

The number of ACS lines $N_a$ (see Figures \ref{single_shot_acs} and 
\ref{pe_terms}) is usually significantly smaller than the nominal size $N_s$ 
of the sampling matrix in the PE direction. For non-EPI imaging sequences the 
ACS data and the reduced data can usually be acquired together by adding a 
few additional k-space lines (for ACS) to the sequence. This method of 
collecting ACS data is usually considered unacceptable for EPI time series 
because it would extend the echo train length significantly and require 
unacceptable compensatory adjustments to the image-space spatio-temporal 
sampling, and because it would exacerbate the EPI-inherent effect of geometric 
distortion due to main field inhomogeneity. Therefore, for GRAPPA EPI time 
series the ACS are usually acquired once prior to the time series and a single 
estimation of the synthesis coefficients is done followed by repeated 
calculations of the missing PE lines from the reduced data sets at each 
discrete time point $n$ of the time series.

There is some flexibility in the scheme used to acquire ACS for an EPI time 
series. For $R$-fold acceleration it is possible to use 1-shot to $R$-shot 
(interleaved) ACS trajectories, each with distinct advantages and caveats. For 
example, a rapid estimation of GRAPPA coefficients may be obtained with a 
$1$-shot ACS (see Figure \ref{single_shot_acs}), using $\Delta k$ matched to 
that of the nominal set of PE lines \citep{Porter, Heidemann_Porter}.
This ACS data set should be relatively uncontaminated by sample motion during 
the ACS acquisition itself. However, the $R$-shot trajectory (see Figure 
\ref{pe_terms}) has the advantage of eliminating artifact that results when the
signal dynamics over a $1$-shot ACS sampling trajectory differ significantly 
from that over the reduced data set trajectory, as may occur in the presence 
of main field inhomogeneity \citep{Cheng} or $T_2^*$ signal relaxation. 
\ref{grappa_inhomo} shows, from a mathematical frames perspective, that the 
GRAPPA equations are expected to break down when significant main magnetic 
field inhomogeneity is present unless the ACS data is acquired in a matched 
$R$-shot trajectory. This breakdown of the GRAPPA equations is expected to be 
exacerbated when using strong main fields since main field inhomogeneity 
increases with main field strength and since a large reduction factor is 
often used to control the resulting geometric distortions due to the field 
inhomogeneity. 

\begin{figure}[H]
  \centering
  \includegraphics[scale = 0.8, viewport = 0 10 470 350,clip]{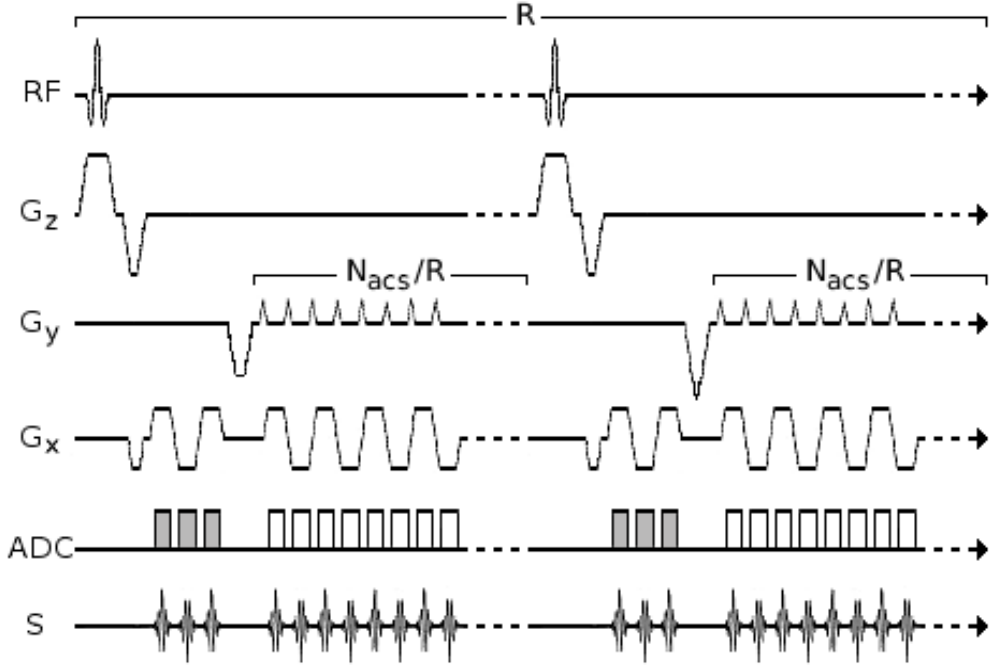}
  \caption{Sequence diagram for the acquisition of one complete set of ACS 
  data. $G_x$, $G_y$ and $G_z$ denote the frequency-encoding, phase-encoding 
  and slicing gradients respectively. The gray-filled ADC regions identify the 
  phase correction data lines which were also used as navigator echoes to 
  detect motion between interleaved ACS segments. Note that the prephasing 
  $G_y$ gradient is incremented for each ACS segment to achieve the 
  interleaving.}
  \label{seq_diag}
\end{figure}

In the remainder of this paper we: (1) Demonstrate the effects of ACS 
trajectories unmatched with respect to field inhomogeneity perturbations
and $T_2^*$ signal relaxation; and (2) Propose and demonstrate a simple method 
by which residual aliasing due to motion between $R$-shot ACS segments may be 
reduced. The phase correction lines (see Figure \ref{seq_diag}) which precede 
each 2D slice in commercial EPI sequences are used as navigator echoes to 
assess whether signal changes due to motion or other perturbations have 
occurred between the segments of a complete ACS data set. Using the data
from these navigator echoes the following metric $M_s$ will be used to 
assess the perturbations between the $R$ segments of the $s^{th}$ complete 
ACS data set:
\begin{eqnarray}
  M_s =  \sqrt{
          \frac 
          {\sum_{m=1}^{N_c} \sum_{n=1}^{N_s} \sum_{j=1}^{R-1} 
            ||S_{sjm}[n] - S_{s,j+1,m}[n]||^2}
          {\sum_{m=1}^{N_c} \sum_{n=1}^{N_s} 
            ||S_{s1m}[n]||^2}}
  \label{mnm_1}
\end{eqnarray}
where $S_{sjm}[n]$, $j$, $m$ and $n$ are the FFT of a line of phase correction 
data, the segment index ($j=1,\ldots, R$), the coil element index and the 
sampling index in the frequency-encoding direction, respectively. We expect 
that complete ACS data sets with the smallest associated $M_s$ will produce 
EPI images with the least residual aliasing. 

\section{Methods}
\label{methods}

A single phantom experiment was performed to demonstrate the artifacts 
encountered when the $1$-shot ACS trajectory is used in the presence of 
significant main field inhomogeneity. Two brain experiments were performed
to show the efficacy of the proposed navigator-based method of acquiring ACS 
with reduced motion contamination.

\subsection{Data Acquisition} 

All data were acquired on a Siemens TIM Trio 3T whole-body scanner with a 
$12$-channel phased array head receiver coil. Brain images were obtained 
from volunteers in accordance with a protocol approved by the institutional
Committee for the Protection of Human Subjects. 

\subsubsection{Phantom Experiment}
In the first experiment we acquired data to compare GRAPPA reconstructed EPI 
images resulting from $R$-shot segmented interleaved 
ACS versus $1$-shot ACS. A structural phantom (The Phantom Lab, Salem, NY, 
USA) was used to assure that the image data was motion-free. An EPI sequence 
was modified in-house to either acquire ACS data for $R=3$ GRAPPA in a $1$-shot 
acquisition or a segmented interleaved $3$-shot acquisition. It should be noted
that Siemens' commercial sequences use a $1$-shot trajectory for $R=2$ and an 
$R$-shot segmented interleaved trajectory for $R > 2$ when acquiring the ACS 
data set.

The relevant imaging parameters for this comparison were as follows: TR = 
$2450$ ms, TE = $33$ ms, nominal matrix size $N_s \times N_s$ = $96 \times 96$,
FOV = $224 \times 224$ mm, number of ACS reference lines $N_a$ = $24$,  
echo-spacing = $1.6$ ms and slice thickness = $3$ mm. This experiment 
facilitated an evaluation of the potential for mismatched field inhomogeneity
to affect residual aliasing. Note that an echo spacing roughly twice as
long as is typical for fMRI was chosen to provide a clear example of the
mismatch problem. We did not set out to evaluate the severity of the artifact
for $1$-shot $R$-fold acceleration per se.

\subsubsection{Brain Experiments}
In a second experiment we demonstrate the proposed 
method of multiple ACS acquisition for eliminating residual aliasing due to 
motion during $R$-shot ACS. Multislice 2D brain images were obtained from 
volunteers using the following imaging parameters: GRAPPA with $R=2$, 
echo-spacing = $0.8$ ms, nominal matrix size $N_s \times N_s$ = $64\times 64$, 
number of phase correction reference scans (which also serve as the navigator 
echoes) = $3$, and TR = $2000$ ms. An EPI sequence was modified in-house 
to acquire $N_{acs} = 10$ complete $R=2$ interleaved ACS data sets 
(see Figure \ref{pe_terms} for a depiction of the analogous $R=3$ case) 
instead of the usual single complete set. Each interleaved ACS EPI segment 
acquired $12$ PE lines for a total of $24$ ACS lines ($N_a = 24$) in a 
complete ACS data set. Following the ACS acquisition 
a single volume of 2D multislice image data (the reduced set) was acquired. 
For fMRI applications a time series of reduced data volumes would be acquired, 
but in this experiment we were only interested in the effects of motion during 
the ACS and therefore a single reduced k-space volume was sufficient.

In order to investigate the various effects of subject motion during 
interleaved ACS, four separate trials were acquired in the second experiment. 
During the acquisition of the first three trials the subject nodded their head 
(approximately 3 degrees) in randomly distributed $4$-second intervals, 
between which the subject tried to remain motionless. During the acquisition 
of the fourth trial the subject tried to remain motionless throughout the 
acquisition of the $10$ complete ACS data sets. The fourth trial was considered 
to be the target, or "best case", data.

In a third experiment we sought to assess the effect of motion contaminated
ACS on the tSNR of a time series EPI acquisition, as would be used for fMRI. 
Once, again the subject was instructed to nod their head during ACS acquisition
after which they were to remain motionless throughout an EPI time series. The 
relevant imaging parameters were $R$ = $2$, number of time series volumes 
$N_{epi}$ = $100$, nominal matrix size $N_s \times N_s$ = $64\times 64$, 
number of ACS reference lines $N_a$ = $24$, number of repetitions of segmented 
interleaved ACS complete data set $N_{acs} = 10$ (as depicted in Figure 
\ref{pe_terms}), number of phase correction reference scans (which also serve 
as the navigator echoes) = $3$, TR = $2000$, TE = $28$ ms, echo-spacing = $0.52$
ms and FOV = $224$ mm.

\subsection{Image Reconstruction} 

Image reconstruction was performed with either commercial Siemens GRAPPA 
algorithms and code or with in-house  GRAPPA reconstruction code which 
implemented "method 1" (a PE-only GRAPPA interpolation with a small number 
of blocks and no sliding window averaging) given in \citet{GRAPPA_comparisons}.

\subsubsection{Phantom Experiment}
For the first experiment image reconstruction was 
done on the scanner with standard Siemens reconstruction code for both the
product and modified sequence. 

\subsubsection{Brain Experiments}
For the second and third experiments the image 
reconstruction was done with in-house GRAPPA code. The EPI sequence acquires 
three phase correction lines prior to each 2D slice of image data which we 
used as navigator echoes. Using Eq. \ref{mnm_1} we calculate $M_s$ associated 
with each phase correction line and then we simply averaged the three $M_s$ 
values to yield the final $M_s$ value for each of the complete ACS data sets 
of a given trial. 

The four trials of the second experiment were processed in the manner of four 
"ACS time series" acquisitions, with each set of the ten interleaved ACS being 
combined with the single reduced EPI data set following each looped ACS.
The $20$ separate ACS segments from ten interleaved $R=2$ ACS 
acquisitions can be grouped into 19 complete ACS data sets using adjacent 
pairs only. We do not expect the order of the interleaved ACS acquisition to 
matter, but we chose to restrict the metric $M_s$ to ACS pairs acquired as 
nearest neighbor pairs in order to minimize motion and scanner drift effects. 
From each complete ACS data set the synthesis coefficients were estimated. The 
missing set of PE lines was then synthesized using the reduced data set of PE 
lines and the synthesis coefficients corresponding to each of the $N_{acs} = 
19$ complete ACS sets. This was done for each of the twelve coil elements, 
after which the data was Fourier transformed to image-space and the images 
from each coil were combined in a final sum-of-squares magnitude image. 

Data from the third experiment was processed in two ways. As in the second 
experiment, $M_s$ was evaluated for 19 nearest neighbor pairs of complete
ACS data sets from the 10 acquired ACS pairs. From these, the ACS pairs with
the highest and lowest $M_s$ were determined and used in the reconstruction
of the same $N_{epi} = 100$ volume time series EPI data, from which tSNR maps
were produced. TSNR images were obtained by calculating the ratio of the 
temporal mean to the temporal standard deviation over the EPI time series 
for each pixel in image space.

\section{Results}
\label{results}

\subsection{Phantom Experiment}
\begin{figure}[H]
  \centering
  \includegraphics[scale = 0.75, viewport = 170 -10 400 150]{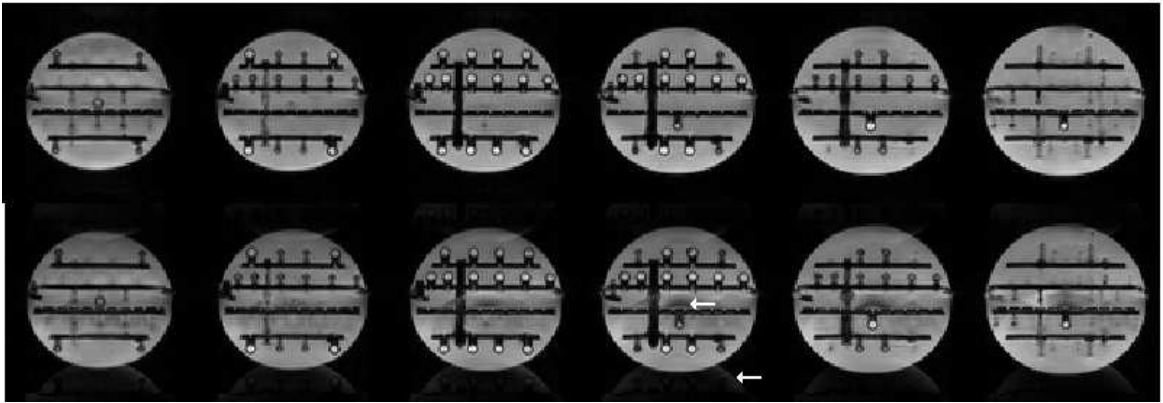}
  \caption{Top row: Five successive slices of $R=3$ GRAPPA images reconstructed
  using a 3-shot interleaved ACS data acquisition. Bottom row: The same set of 
  image slices but using a single-shot ACS data acquisition. Note the residual 
  aliasing (white arrows) which extends throughout the images of the bottom row.
  Exterior to the phantom the residual aliasing appears as misplaced image 
  intensity, while interior to the phantom the residual aliasing appears as 
  both increased and decreased local image intensity.}
  \label{res_alias}
\end{figure}

The bottom of row of Figure \ref{res_alias} shows, 
for the case of $R=3$, an example of the residual aliasing artifact that may 
result when $1$-shot ACS data is used to autocalibrate the synthesis 
coefficients. Such artifact is clearly undesirable and is much reduced by
using $3$-shot segmented interleaved ACS instead, as shown in the top row of
Figure \ref{res_alias}. We did not seek to quantify the extent of residual
aliasing because the amount depends on $R$, the main field inhomogeneity (which
itself may depend upon main field strength and the heterogeneity of the
magnetic susceptibility of the imaged object), echo-spacing time and the FOV. 
The effects of field inhomogeneity upon the GRAPPA equations are considered in 
\ref{grappa_inhomo}.

\subsection{Brain Experiments}
\begin{figure}[H]
  \centering
  \includegraphics[scale=0.34, viewport = 550 -50 750 860]{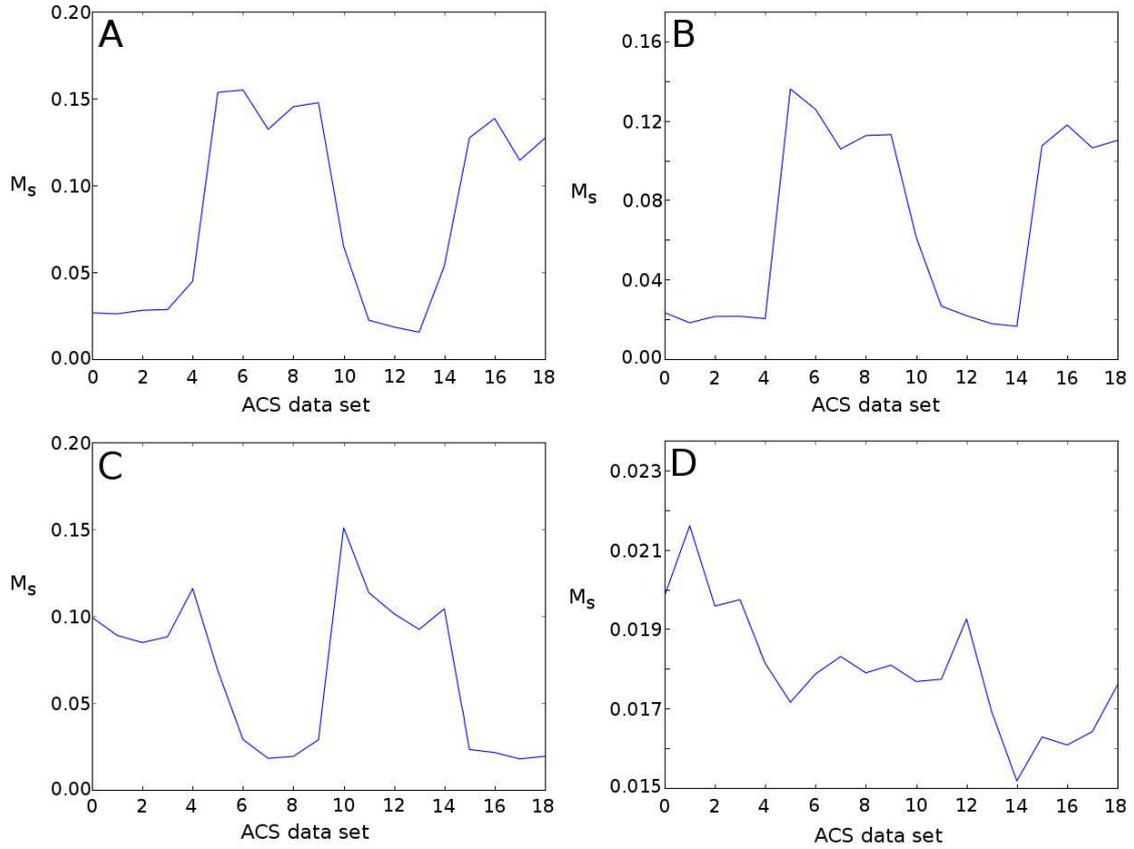}
  \caption{Plots of the perturbation metric $M_s$ as a function of the $n^{th}$
  complete ACS data set. Plots A through D correspond respectively to rows 1 
  through 4 of Figure \ref{acs_results}. Note that the scale of the ordinate 
  varies between the four plots.}
  \label{metric}
\end{figure}

\begin{figure}[H]
  \centering
  \includegraphics[scale=1.0,  viewport = 160 475 330 700]{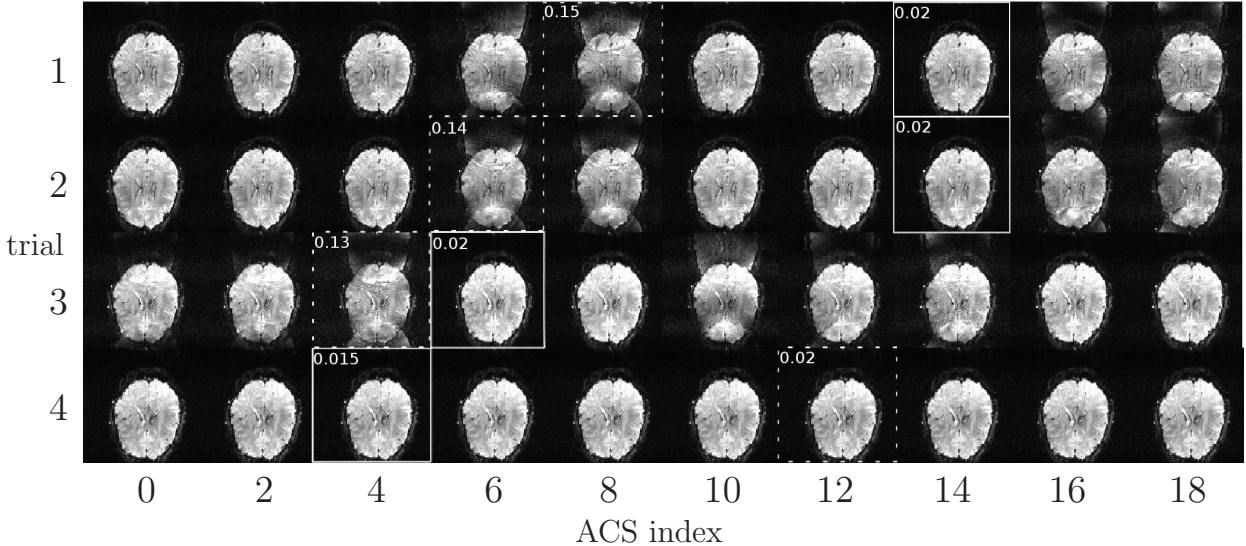}
  \caption{Images reconstructed from ACS data sets during which subject motion 
  was introduced. Image rows 1 through 4 correspond to four different trials. 
  During the acquisition of trials 1-3 the subject was instructed to move 
  during ACS acquisition. During trial 4 the subject was instructed to remain 
  motionless during the ACS acquisition. For brevity only the $10$ images 
  associated with the even valued complete ACS data sets are shown rather than    
  all $19$. For each trial the image corresponding to the highest or lowest 
  value of the ACS perturbation metric $M_s$ is indicated by a dashed or solid 
  bounding box respectively. The value of $M_s$ is given within the bounding 
  box.}
  \label{acs_results}
\end{figure}

Figures \ref{metric} and \ref{acs_results} show the results for a GRAPPA-EPI 
$R=2$ acquisition acquired in the second experiment. Figure \ref{metric} shows 
plots of the perturbation metric $M_s$ associated with each of the 19 complete 
ACS data sets for the four motion trials. Although the perturbation metric in 
the first three trials attains values nearly an order of magnitude greater 
than those of the fourth (no intentional motion) trial, it is clear that low 
$M_s$ sets of ACS data are available in all cases. 

Figure \ref{acs_results} 
shows the EPI images associated with successive complete ACS data sets. For 
brevity only the $10$ images associated with the even valued complete ACS data 
sets are shown, rather than all $19$. Image rows 1 through 4 correspond to the 
four different trials during which varying amounts of motion were introduced. 
Image row 4 was acquired as the subject tried to remain motionless throughout 
ACS acquisition. For each of the runs, the dashed boxes enclose the GRAPPA 
reconstructed images for which $M_s$ was highest and the solid boxes enclose 
the images for which $M_s$ was lowest. In each of the enclosed images the 
value of $M_s$ is given in the upper left corner. Figures \ref{metric}  and 
\ref{acs_results} show a clear correlation between the perturbation metric 
$M_s$ and the visually apparent residual aliasing.

Figure \ref{tsnr_results} shows two mosaics of tSNR images for the same 100
volume EPI time series data acquired in the third experiment. The top mosaic 
is associated with the largest valued perturbation metric during the ACS 
while the bottom mosaic is associated with the smallest valued perturbation
metric. 

\noindent metric. Both images were constructed using the same gray-value scale. Visual 
inspection of these images shows significant improvement in tSNR of the EPI 
time series when the ACS data associated with the smallest valued perturbation 
metric is used in the calculation of the GRAPPA synthesis coefficients. For 
reference note that for the $5 \times 5$ pixel region shown in Figure 
\ref{tsnr_results} the ratio of the mean tSNR for the largest valued metric 
case to the mean tSNR for smallest valued metric case is 
$tSNR_{smallest}/tSNR_{largest} = 1.54$,  the improvement in tSNR being due
to the proper localization of the signal in image space.

\section{Discussion}
\label{discuss}

In this work we demonstrated the effect of field inhomgeneity for one example 
situation to highlight the severity of artifacts that can result from $1$-shot
ACS. (The mismatch phenomenon is treated in depth in \ref{grappa_inhomo}.) 
With increasing $R$, field inhomogeneity or EPI 
echo-spacing time the effects of main field inhomogeneity on $1$-shot ACS data 
acquisition are expected to become increasingly pronounced, suggesting a 
growing need for interleaved ACS data acquisition when acquiring ACS data for 
GRAPPA-accelerated EPI at 3T and above. 

It is important to not substitute one source of error for another whenever 
possible. Therefore we have developed and demonstrated in this paper a means 
of acquiring segmented interleaved ACS data which greatly diminishes ACS motion
contamination. Acquiring redundant ACS prior to \nolinebreak[4] an

\par\vfill\break 
\advance\voffset by -3cm 

\begin{figure}[H]
  \centering
  \includegraphics[scale=1.2]{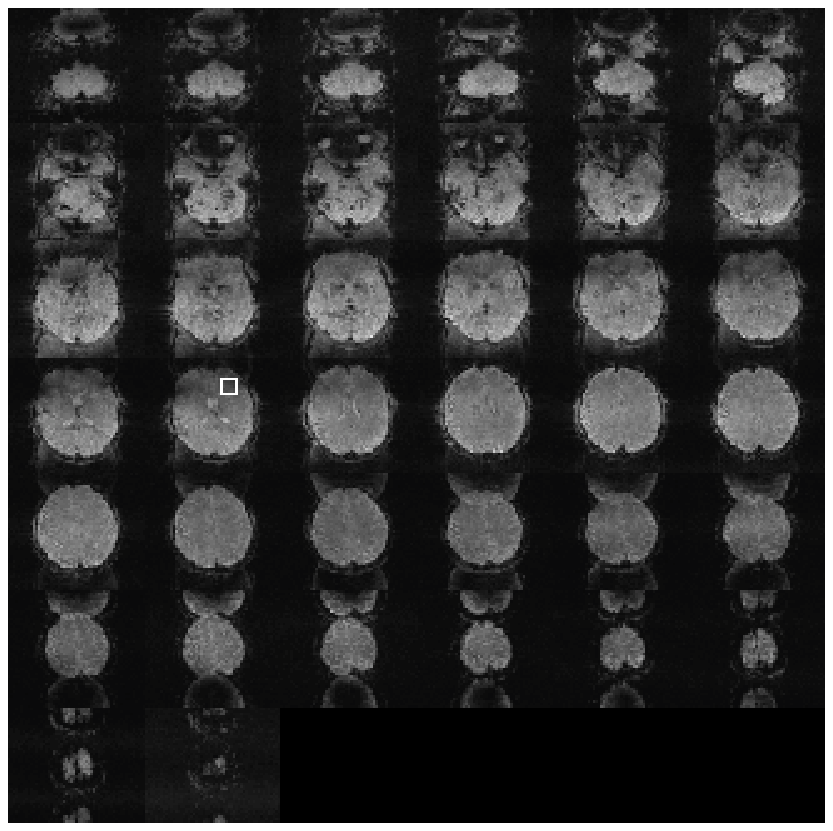} \\
  \vspace{-15pt}
  \includegraphics[scale=1.2]{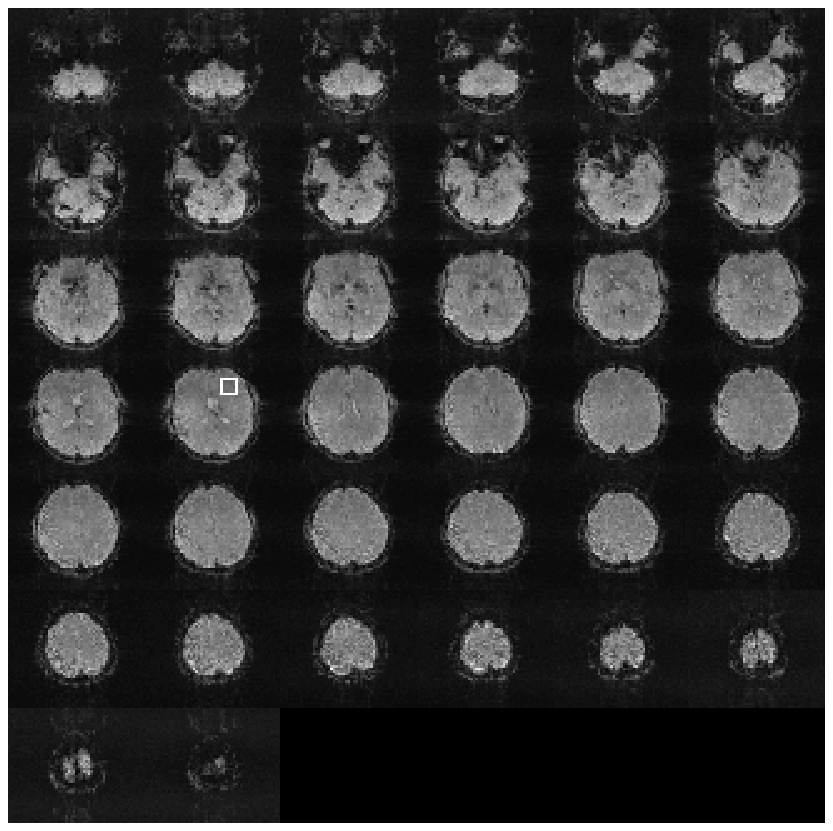}
  \caption{2D multislice mosaics of tSNR images for the EPI time series. Top:
  Time series reconstructed by using the complete ACS data set with the largest
  valued perturbation metric. Bottom: Time series reconstructed by using the 
  complete ACS data set with the smallest valued perturbation metric. Note that
  the same 100 volumes of accelerated ($R=2$) time series EPI data were used 
  here, and only the ACS used to compute the synthesis coefficients differs 
  between the two sets of images. For reference, the ratio of the mean tSNR, 
  over the $5x5$ pixel region outlined in white, for smallest and largest 
  valued $M_s$ is $tSNR_{smallest}/tSNR_{largest} = 1.54$.}
  \label{tsnr_results}
\end{figure}

\par\vfill\break 
\advance\voffset by 3cm 

\noindent accelerated EPI time series allows the use of navigator echoes, 
conveniently already part of a product EPI pulse sequence, as a means of 
assessing motion corruption of GRAPPA ACS data. We have demonstrated that 
it is possible to retrieve the lowest motion-contaminated ACS data from a 
series of ACS data sets. Moreover we have demonstrated that motion during 
ACS has important consequences with respect to image artifact and time series 
tSNR. It is expected that this simple modification will greatly enhance the 
robustness of GRAPPA-accelerated EPI for fMRI, where subject motion during 
the initial part of the acquisition - the interleaved ACS - can render an 
entire time series worthless in a worst-case scenario, or with compromised 
tSNR as demonstrated in Figure \ref{tsnr_results}.

In this work we acquired a fixed number of complete ACS data sets but the 
method could be further extended by looping on the ACS acquisition until
the metric $M_s$ meets a prior criterion, thereby assuring that acceptable 
GRAPPA coefficients will be obtained in the presence of protracted motion
during ACS acquisition. Figures \ref{metric}  and \ref{acs_results} show that 
it should be possible to establish such a criterion for the metric $M_s$, 
although it would necessitate an automated way to advance the pulse program 
to the time series acquisition phase.

One limitation of the present method is a potential directional dependence 
of the navigator echoes. Since the navigators are projections of a 2D image 
slice (multiplied by a receive coil field) onto the frequency encoding axis, 
this method should be relatively insensitive to motion that is in the 
PE direction only. The nonuniformity of the receive coil field will introduce 
some sensitivity to motion that is in the PE direction only but if enhanced 
sensitivity to this type of motion is needed then a second navigator, 
spatially orthogonal to the first, could be used.  

The work reported in this paper stands in contrast to the TGRAPPA EPI time
series method investigated by Cheng \citep{Cheng}. In the TGRAPPA EPI times 
series method the effects of motion upon autocalibration are addressed by 
updating the synthesis coefficients at each point in the time series. In our 
work we take the perspective that a single calculation of the synthesis 
coeffcients will generate the greatest stability in the synthesis of the 
missing PE lines. We find support for this perspective in Cheng's TGRAPPA
work where he noted that in the absence of significant motion the use of 
frequently updated synthesis coefficients leads to a reduction of tSNR 
compared to the usual single calculation of synthesis coefficients. Indeed, in 
separate work, Cheng has shown recently \citep{Cheng2} that the temporal noise 
properties of GRAPPA EPI time series can be reduced via a single set of ACS 
for multiple time series acquisistions, rather than one ACS per acquisition. 
This increases still further the importance to acquire a motion-free ACS. 
Thus, the method presented here shows that it is possible to reduce the 
effects of motion during ACS acquisition while obtaining the desirable 
properties of the $R$-shot interleaved ACS, i.e. matched distortion 
characteristics for the ACS and the accelerated time series data. This should 
be of special interest in fMRI where motion is especially problematic.

From our perspective any motion during the time series and following the ACS 
leads to a spatial undersampling of the receive coil fields - the head may 
have moved into a region of the array from which no signal (or signal of low 
SNR) originated during the ACS. We see this spatial undersampling as a separate
issue to be addressed in future work. It is a more involved problem than the 
simple method presented here that permits the acquisition of 
ACS uncontaminated by motion.

\appendix

\section{GRAPPA and Field Inhomogeneity}
\label{grappa_inhomo}

We assume that there are $N_c$ receive coils in the receive array. For the
reduced data set the signal from the nth receiver coil, after Fourier 
transform in the frequency-encoding $x$-direction and phase correction 
(Nyquist ghost correction) \citep{RFAM}, may be written as
\begin{eqnarray}
  s_n(x,mR) = \langle \rho, g_{nm} \rangle 
  = \int \rho(x,y) g^*_{nm}(x,y) dy
  \label{sig4}
\end{eqnarray}
where
\begin{eqnarray}
  g_{nm}(x,y) = g_n^*(x,y) e^{i 2\pi m R\Delta k y} e^{-i m \tau_e \phi(x,y)}
  = E^m  g_n^*(x,y). 
  \label{sig5}
\end{eqnarray}
and where the definition of $E$ should be obvious, $g_n(x,y)$ is the receiver 
coil field (coil sensitivity), $R$ (an integer) is the reduction factor, 
$\phi(x,y)$ is the local field inhomogeneity offset of the main field, 
$\tau_e$ is the echo-spacing time and $m \in \mathbb Z$. Note that 
$\rho(x,y)$ is assumed to be an image that is compactly supported on $|y|, |x| 
< 1/\Delta k$ and that the PE direction, $y$, will be the direction of 
acceleration. Also note that $s_n$, $g_n$ and $\phi$ will depend 
upon $z$ as well but we have omitted explicitly writing this dependence.
Note that although $\langle \rho, g_{nm} \rangle$ is a function of $x$
we will suppress the $x$-dependence for the sake of an economy of symbols. 

When the ACS data are acquired in $R$ interleaved segments the signal in the 
$n^{th}$ receiver coil will be
\begin{eqnarray}
  s^R_n(x,mR+p) = \langle \rho, g^R_{nm} \rangle 
  = \int \rho(x,y) g^{R*}_{nm}(x,y) dy
  \label{sig4.1}
\end{eqnarray}
where $p=0, \ldots, R - 1$ and
\begin{eqnarray}
  g^R_{nm}(x,y) = g_n^*(x,y) e^{i 2\pi (m R + p)\Delta k y}
  e^{-i m \tau_e \phi(x,y)}
  = E^m E_{Ro}^p g_n^*(x,y). 
  \label{sig5.1}
\end{eqnarray}
The superscript $R$ denotes the R-shot trajectory for the complete ACS data 
set. When the complete ACS data set is acquired in one segment ($1$-shot 
trajectory) then the signal may be written as 
\begin{eqnarray}
  s^1_n(x,mR+p) = \langle \rho, g^1_{nm} \rangle 
  = \int \rho(x,y) g^{1*}_{nm}(x,y) dy
  \label{sig4.11}
\end{eqnarray}
where
\begin{eqnarray}
  g^1_{nm}(x,y) = g_n^*(x,y) e^{i 2\pi (m R + p)\Delta k y}
  e^{-i (m R + p) \tau_e \phi(x,y)}
  = E_1^m E_{1o}^p g_n^*(x,y). 
  \label{sig5.11}
\end{eqnarray}
and where the definition of the modulation operators $E_R$, $E_{Ro}$ $E_1$
and $E_{1o}$ should be obvious. The superscript $1$ denotes the $1$-shot 
trajectory for the complete ACS data set.

For a suitable $g_n(x,y)$ and $R$ the set $\lbrace g_{nm} \rbrace$, where 
$n = 0, \ldots, N_c - 1$ and $m \in \mathbb Z$, will form a frame \citep{Ole}. 
Since $\lbrace g_{nm} \rbrace$ (reduction factor $R > 1$) is assumed to be 
a frame we can write
\begin{eqnarray}
  \rho = \sum_{m'=-\infty}^\infty \sum_{n'=1}^{N_c} 
  \langle \rho, g_{m', n'} \rangle 
  S^{-1} g_{n'm'}
  \label{sig6a}
\end{eqnarray}
where $S$ is the frame operator for the frame $\lbrace g_{nm} \rbrace$ and 
where $\lbrace S^{-1} g_{nm} \rbrace$ is the canonical dual frame to the 
frame $\lbrace g_{nm} \rbrace$ \citep{Ole}. Note that the dual frame depends 
upon $x$ although we will usually not indicate this explicitly. Note also 
that the canonical dual frame is only one of the possible and non-unique dual 
frames for which Equation (\ref{sig6a}) may be written. The canonical dual 
frame is a special dual frame in that it has the property that the $\langle 
\rho, S^{-1} g_{m', n'} \rangle$ have a minimal $l^2$-norm among all possible 
dual frames. 

If the inverse frame operator could be found then equation (\ref{sig6a}) 
would form the fundamental means of obtaining $\rho$. In practice obtaining 
the inverse frame operator is mathematically difficult to obtain and this is 
at least partly due to the lack of an explicit knowledge of the coil 
sensitivities $g_n(x,y)$ (or the field inhomogeneity which may perturb the 
frame). Autocalibration methods provide an alternative to obtaining estimates 
of the dual frame and instead focus on determining the coefficients that 
transform one frame to another. We now take a closer look at the GRAPPA 
autocalibration method and how main-field inhomogeneity may affect the
method. 

Performing an analysis of $\rho$ given by equation (\ref{sig6a}) with respect 
to the frame $\lbrace g^R_{nm} \rbrace$ we obtain
\begin{eqnarray}
  \langle \rho, g^R_{nm} \rangle 
  = \sum_{m'=-\infty}^\infty
  \sum_{n' = 1}^{N_c} \langle \rho, g_{n'm'} \rangle 
  \langle 
    S^{-1} g_{n'm'}, g^R_{nm} 
  \rangle.  
  \label{sig9}
\end{eqnarray}
which we may write as
\begin{eqnarray}
  \langle \rho, g^R_{nm} \rangle 
  = \sum_{m'=-\infty}^\infty
  \sum_{n' = 1}^{N_c} \langle \rho, g_{n'm'} \rangle 
  \langle 
    S^{-1}  E^{m'} g_{n'}^*, 
    E_R^m E_{Ro}^p g_{n} 
  \rangle.  
  \label{sig9.1}
\end{eqnarray}
Since the frame operator $S$, and therefore its inverse, commutes with the 
modulation operator $E^m$ \citep{Ole} then equation (\ref{sig9}) may be written
as 
\begin{eqnarray}
  \langle \rho, g^R_{nm} \rangle
  = \sum_{m'=-\infty}^\infty \sum_{n' = 1}^{N_c}  
  \langle \rho, g_{n'm'} \rangle 
  \langle 
    E^{m'-m} S^{-1} g_{n'}^*,  E_{Ro}^p g_n 
  \rangle
  . 
  \label{sig17}
\end{eqnarray}
or according to equations (\ref{sig4}) and (\ref{sig4.1}) 
\begin{eqnarray}
  s^R_n(x,mR + p)
  =  \sum_{m'=-\infty}^\infty \sum_{n' = 1}^{N_c}  
  s_{n'}(x,m'R) \; c(m'- m, n', n, p, x)
  \label{sig17.3}
\end{eqnarray}
where we have defined the synthesis coefficients 
\begin{eqnarray}
  c(m'- m, n', n, p, x) = 
  \langle 
    E^{m'-m} S^{-1} g_{n'}^*,  E_{Ro}^p g_n 
  \rangle
  \label{sig17.3b}
\end{eqnarray}
Notice that the $c$ depend upon the difference $m - m'$ only. This particular 
dependence upon $m$ and $m'$ allows us, through a change of indices and a 
truncation of the sum over $m'$ (which must occur in a any practical setting), 
to write 
\begin{eqnarray}
  s^R_n(x,mR + p)
  = \sum_{m'=b_1}^{b_2} \sum_{n' = 1}^{N_c} 
  s_{n'}(x,mR + m'R) \; c(m', n', n, p, x) 
  \label{sig17.4}
\end{eqnarray}
which is equivalent to the form of the GRAPPA equations given by Equation 
(\ref{sig0b}). The quantity $B=b_2 - b_1$ is often referred to as the {\it
block size}.

When the GRAPPA equations and the complete ACS data set are used to solve 
for the coefficients $c(m', n', n, p, x)$ it is usually the case that the 
number of unknown coefficients $c(m', n', n, p, x)$ exceeds the number of 
equations. When the number of ACS lines $N_a$ is an interger multiple of the
reduction factor $R$ then there will be $N_c N_a (R-1)/R$ equations and
$B N_c N_a /R$ unknown coefficients $c(m', n', n, p, x)$ at each point $x$. 
For a typical example of $N_a = 24$, $N_c = 12$ and $R=2$ the number of 
equations would be $144$ and the number of unknowns is $144B$. This presents 
a problem with respect to meeting necessary (but not sufficient) requirements 
for a unique solution set of the $c(m', n', n, p, x)$ - number of equations 
must be greater than or equal to the number of unknowns. To meet the necessary 
requirement it is usually assumed that the $c(m', n', n,p, x)$ vary little 
with respect to $x$ in the neighborhood of any $x$ in the FOV but that the 
signal does vary significantly with $x$. This allows one to use the ACS data 
at neighboring $x$ (at least $(d_2-d_1)/(R-1)$ are needed) to locally determine
the $c(m', n', n, p, x)$. Such an assumption implies that $g_n$, $S^{-1}$ and 
$\phi$ vary little about any $x$ in the FOV. 

When the 1-shot trajectory is used to sample the complete ACS data set then,
following steps similar to those given above, we can write
\begin{eqnarray}
  s^1_n(x,mR + p)
  = \sum_{m'=-\infty}^\infty \sum_{n' = 1}^{N_c}  
  s_{n'}(x,m'R) 
  \langle 
    E^{m'} E_1^{-m} S^{-1} g_{n'}^*,  E_{1o}^p g_n 
  \rangle
  . 
  \label{sig18}
\end{eqnarray}
Since $E \ne E_1$ it is not possible to cast Equation (\ref{sig18}) into the 
form of Equation (\ref{sig17.3}) because the indices $m$ and $m'$ do not
appear as a difference only. Therefore, except when $\phi(x,y) = 0$, it is 
necessary that the ACS data be collected over an R-shot trajectory in order 
for GRAPPA equations to strictly apply. As $R$ or $\phi(x,y)$ increases in 
magnitude the departure from the GRAPPA equations will increase. We can 
therefore expect the artifacts resulting from a 1-shot ACS calibration scan 
to increase with main field strength where susceptibility effects will be 
exacerbated and increasing reduction factors may be applied to combat the 
resulting distortions due to field inhomogeneity.

\bibliographystyle{elsarticle-harv}
\bibliography{mag_res}

\end{document}